\documentclass[12pt]{article}
\usepackage{graphicx}
\usepackage[top=3cm, bottom=2cm, left=3cm, right=3cm]{geometry}

\title{Response of atmospheric ground level temperatures to changes in the 
total solar irradiance}
\author{$^*$A. D. Erlykin $^{1,2}$, and A. W. Wolfendale $^{2}$\\
$(1)$ P.N.Lebedev Physical Institute, Leninsky prosp., Moscow, Russia\\
$(2)$ Department of Physics, Durham University, Durham, UK
\footnote{corresponding author: tel.+74991358737, e-mail: erlykin@sci.lebedev.
ru}}
\begin{document}
\maketitle

\begin{abstract}
The attribution of part of `global warming' to changes in the total solar irradiance
(TSI) is an important topic which is not, yet, fully understood.
Here, we examine the TSI-induced temperature (T) changes on a variety of time scales,
from one day to centuries and beyond, using a variety of assumptions. Also considered 
is the latitude variation of the T-TSI
correlations, where it appears that over most of the globe there is a small
increase in the sensitivity of temperature to TSI with time.
It is found that the mean global sensitivity $\alpha$ measured in $K (Wm^{-2})^{-1}$
varies from about 0.003 for 1 day, via 0.05 for 11-years to $\sim$0.2 for decades to
centuries. We conclude that mean global temperature changes related to TSI are not 
significant from 1975 onwards.

Before 1975, when anthropogenic gases were  less important, many of the temperature 
changes can be attributed to TSI variations. Over much longer periods of time, Kyear to
 Myear, the TSI changes are more efficient still, $\alpha$ increasing to about 0.5. 
Since 1975 the changes in mean global temperature are not due to TSI changes, but, 
rather to the increasing atmospheric $CO_2$ content.
\end{abstract}


\section{Introduction}
The role of changes in the Total Solar Irradiance (TSI) in the global warming debate is
 finite but uncertain; indeed, the latest IPCC Report (IPCC 2014a) quotes estimates for
 the radiative forcing from 1750 to the present of 0.0 to 0.1 Wm$^{-2}$ i.e. a very 
 large range. It also refers to the fact that the value is model and time 
scale dependent. It is true that the TSI averaged over the Earth is approximately 
340Wm$^{-2}$ (as distinct from the value of $\sim 1361 Wm^{-2}$ for incidence along the 
Sun-Earth direction) and the changes are only a fraction of one\%, nevertheless, there 
are two main reasons for a detailed analysis, as follows.
\begin{itemize}
\item[(i)] Since the effect of global warming will be severe, it is imperative to 
understand every facet. The interaction of many factors affecting climate - with the 
attendant positive and negative feedback effects - can give rise to counter-intuitive 
correlations which need to be understood.
\item[(ii)] An interesting aspect of the TSI variations concerns the apparent cloud 
cover, sunspot number (SSN) correlation, which some authors (e.g. Marsh and Svensmark 
2000) have attributed to cosmic rays causing atmospheric nuclei, which in turn form 
cloud droplets, the intensity of which is well 
correlated with SSN. However, others, including ourselves (Erlykin et al. 2009) have 
presented evidence favouring TSI which is, of course, correlated with SSN. Since 
changes in cloud cover are intimately connected with climate change it is important to 
understand the effect of TSI changes.
\end{itemize} 

We define the correlation (or, more strictly, the 'sensitivity') of (ground) 
temperature T with TSI as $\alpha$, in units of
$K(Wm^{-2})^{-1}$; linearity of changes of T with changes of TSI is assumed (see 
\S2.1). The different values of $\alpha$ by different workers have been considered by 
(IPCC 2014a; Erlykin and Wolfendale 2010; Lockwood 2007) and these works have not been 
analysed further. However, the values of $\alpha$ are included in the final Figure for 
comparison. In the present work we make another examination and study latitude 
variations and temporal range effects, i.e. the extent to which induced temperatures 
depend on the length of time over which the increased radiation occurs. The former is 
because the latitude variation might help to explain the mechanism by which the changes
 in TSI are converted to temperature. The role of $CO_2$ in climate change is examined
 and the year in which $CO_2$ rather than TSI starts to dominate. In future, we will 
assume that '$CO_2$' represents all anthropogenic gasses.

The changes caused by ENSO (ElNino Southern Oscillation) and PDO (Pacific Decadal 
Oscillation ) are ignored because these are only partly Sun-induced (~obviously~) but 
consider their effect in \S3.9. Similarly, we disregard volcanoes and aerosol effects 
at our level of accuracy. It is not impossible that the effects, other than TSI 
changes, contribute to the apparent TSI - Temperature correlation and in this case our 
$\alpha$-values represent upper limits. This aspect is considered later.

Inevitably, in this paper there will be some overlap with previous work. However, in 
view of the importance of the topic - and the new aspects - the work is considered to 
be justified.
\section{Solar Variations}
\subsection{The basic data}
We start by considering the basic data. Figure 1 gives the time series for TSI (TSI 
2014) adopted by us. This relates to the total solar irradiance incident upon the 
Earth's atmosphere over all wavelengths. The errors, random and systematic, are not 
known but they are unlikely to be greater than $0.07 Wm^{-2}$ (i.e. $\sim 0.5$\% of the
 total, this being the average dispersion from year to year. We are mindful of problems
 of intercalibration of satellite data (Zacharias 2014) but much of our analysis 
relates to data over 11-year intervals only. The dependencies of 
`temperature' on time for two, important, latitude ranges are also shown (Zonal 
Temperatures 2014).
\begin{figure}[htb]
\begin{center}
\includegraphics[width=16cm,height=15cm]{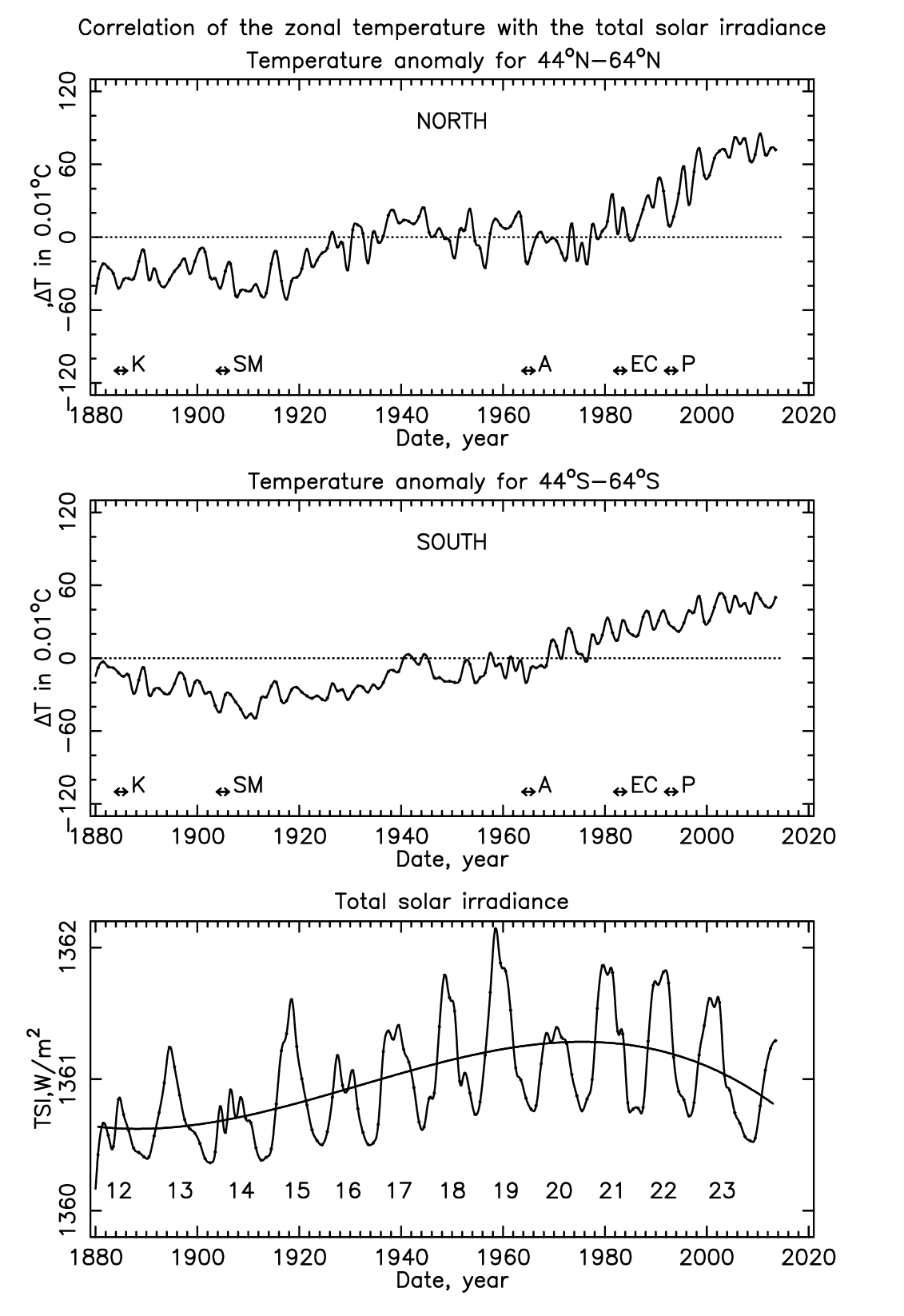}
\end{center}
\caption{\footnotesize Basic data for the correlation of the zonal temperatures [5]
with the total solar irradiance (TSI) [4] for the latitude ranges indicated.
A smooth 3-degree polynomial fit has been made by us to the TSI data.
The significant volcano time ranges are indicated. Key: K: Krakatoa; SM: Santa Maria;
A: Aurang; EC: El Chichon; P: Pinatubo.}
\label{fig:fig1}
\end{figure}

It will be noted immediately that the year by year dispersion of the temperature is
much bigger in the
North than in the South. After 1975 the temporal trends of the temperatures differ
between North and South. This year (1975) is also where the TSI starts to fall.

Inspection of Figure 1 shows that for the Northern range $\Delta$T is increasing slowly
 with time. It means that part of the $\alpha$-values will arise from this cause. If the increase in $\Delta$T were due to solar effects alone, the derived $\alpha$-value would be valid; however, if there is also another cause (the increase in CO$_2$ density is significant) then it will be an upper limit.

The Southern results (Fig. 1, 1880-1975) are superior, in fact, because, as will be
seen from Figure 1, the $\Delta$T values do not have an overall upward trend, they are
the same in 1880 and 1975. Thus, the sensitivity derived from the data ( see Figure 2 )
  should be reliable for the $44^{\circ}-64^{\circ}$ range.

\subsection{The overall $\Delta$T, TSI correlation from 1880}
Figure 2 shows the dependence of $\Delta$T on TSI for data for each full hemisphere and
 for two periods: 1880-1975 and 1976-2013. The values of $\alpha$ and correlation
coefficient {\em r} are shown in the Figure.
\begin{figure}[htb]
\begin{center}
\includegraphics[width=11.5cm,height=16cm,angle=-90]{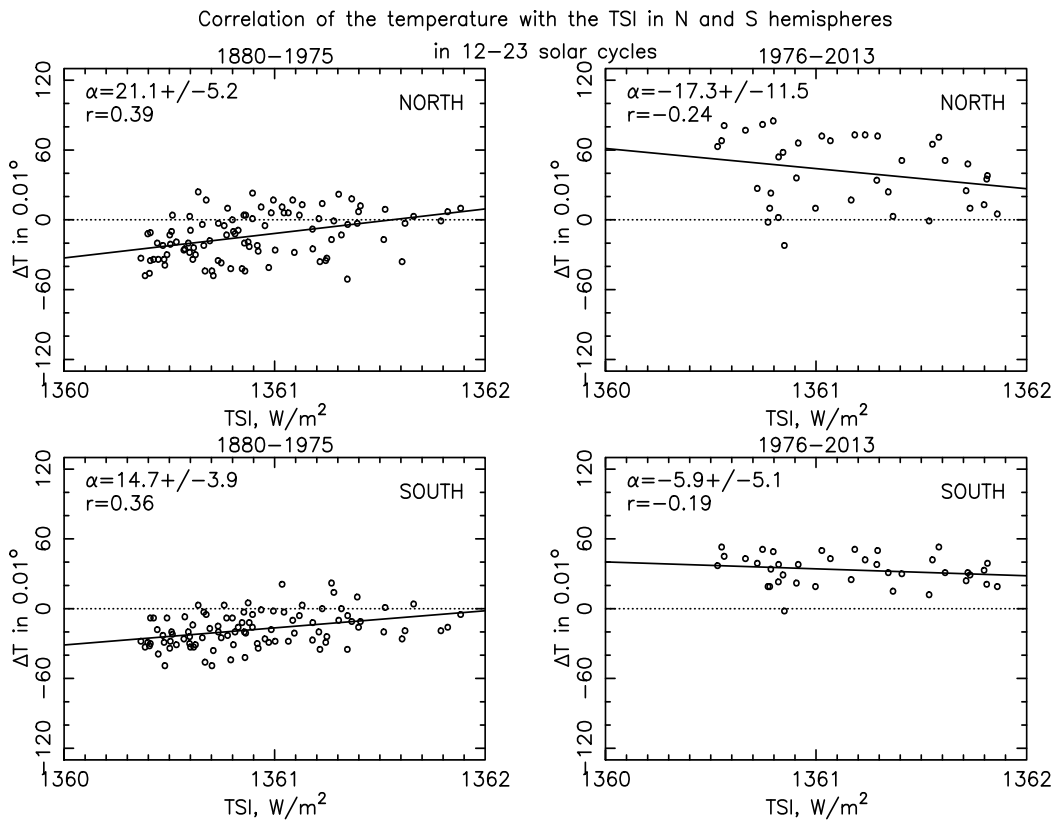}
\end{center}
\caption{\footnotesize Correlation of the temperature anomaly $\Delta$T with the TSI
for the two hemispheres and Solar Cycles 12-20 and 21-23 inclusive. The $\alpha$-values
 are given in the boxes.}
\label{fig:fig2}
\end{figure}

Three features should be noted. \\
(i) It is evident that 1975 marked the transition. Positive $\alpha$ values for
1880 - 1975 change sign and become negative after 1975. It means that after 1975 the
temperature anomaly $\Delta$T still goes up despite the fact that TSI goes down.
This trend confirms the behaviour seen in Figure 1 for 44-64$^\circ$ latitude bands. \\
(ii) The sensitivity, $\alpha$, of $\Delta$T to TSI is higher in the Northern 
hemisphere than in the Southern hemisphere. \\
(iii) Fluctuations of $\alpha$ measured for each solar cycle and latitude band, which
are seen as a spread of points around the best fit line, are larger in the North than
in the South.

All the above characteristics, pre-1975 - can be understood, in principle by the manner
 in which the 'ocean fraction' differs from one latitude band to another: the higher 
the ocean fraction, the higher the thermal inertia and consequent depression of the 
 (short term) temperature variations.

\subsection{Previous Decades}
Although the accuracy of Mean global temperature estimates, and those of TSI are worse
in the past than contemporary values, they are of value. Particularly useful are the
plots of $\Delta$T and $\Delta$(TSI) versus time from (Lean 2004) which 
relate to 1600 to the present (from which we use 1600 - 1800). This range is 
complementary to our own studies of the period from 1900 to the present. Here, the 
sensitivity is $\alpha$ = (0.24$\pm$0.05) $K(Wm^{-2})^{-1}$. $\Delta$T follows 
$\Delta$(TSI) faithfully, including the `Little Ice Age' centered on about 1660, in 
apparent contradictions to the findings of (Foukal et al. 2012).

Another study has been made by (Scarfetta and West 2007, from which $\alpha$ can be 
derived for two forms relating to the TSI. These give $\alpha$ = 0.21 and 
$\alpha$ = 0.68 in the usual units. The former, smaller, value of $\Delta$(TSI) over 
the Little Ice Age is from (Wang et al. 2005) and relates to data reported 5 years 
later. Thus, the value $\alpha$ = 0.21 is preferred. It should be remarked that these 
values do not suffer from the problem of $<T>$ changes due to CO$_2$ increases.

\section{The search for single solar cycle correlations and for longer periods}
\subsection{The latitudinal distribution of $\alpha$ for single Solar Cycles.}
Figure 3 shows the $\alpha$-values versus Cycle No. for each latitude band.
\begin{figure}[htb]
\begin{center}
\includegraphics[width=16cm,height=13cm]{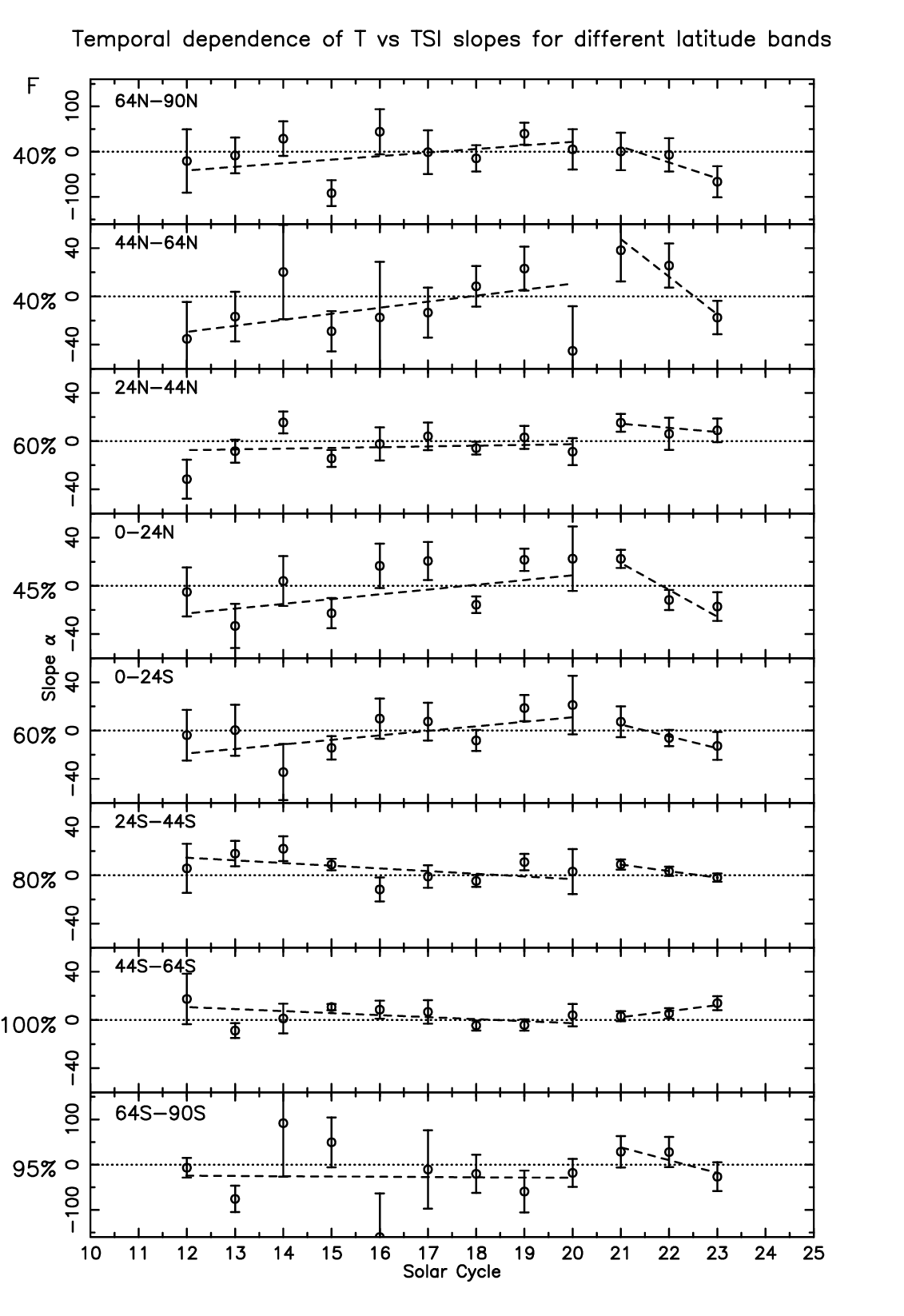}
\end{center}
\caption{\footnotesize Values of $\alpha$ versus Solar Cycle No for each latitude
range. The straight lines are linear fits to the data both before and after Cycle 21.
This Cycle is chosen because it is nearest to the peak of the TSI (see Fig.1) and
also where the density of CO$_2$ starts to increase significantly. Note that the
vertical scale for polar regions 64N-90N and 64S-90S is different from the other
latitude bands. F is the 'ocean fraction'.}
\label{fig:fig3}
\end{figure}
A number of features are worthy of note.
\begin{itemize}
\item[i.] There is a systematic increase of $\alpha$ with time up to 1975 for most
latitude bands (5 out of 8), which reverses after 1975.
\item[ii.] There is no sign of an increased value of $\alpha$ for near-equatorial
latitudes, although absolute values of the temperature are the highest at the equator
 due to geometrical effects: the average solar intensity at Earth being clearly a
diminishing function of latitude.
\item[iii.] Inspection of the TSI as a function of time (Figure 1) shows that the TSI
was highest for Cycle 19. Cycle 19 (Figure 4) shows a consistently high value of
$\alpha$, at least for latitude ranges from 64 - 90N down to 24 - 44S.
\item[iv.] There is a general agreement from latitude to latitude for each Cycle No, 
within the statistical errors where the best lines are taken as the 'datums'. 
Specifically, in terms of the individual errors, the differences from the best line are
 38\% (32\%) above 1$\sigma$ and 5\% (5\%) above 2$\sigma$, where $\sigma$ is the 
standard deviation and the values in brackets are the percentages expected for a 
Gaussian distribution of errors.
\end{itemize}

An important point is that the results suggest that variations within limited regions 
of the Earth (latitude bands) - particularly in the Northern Hemisphere - are 
representative of the whole Earth. For example, large European areas can be so 
representative. Inspection of the Solar Cycles which experienced major volcanoes 
(Figure 1) in Figure 5, show virtually no difference from the other Cycles.

\begin{figure}[h]
\begin{center}
\includegraphics[width=8cm,height=16cm,angle=-90]{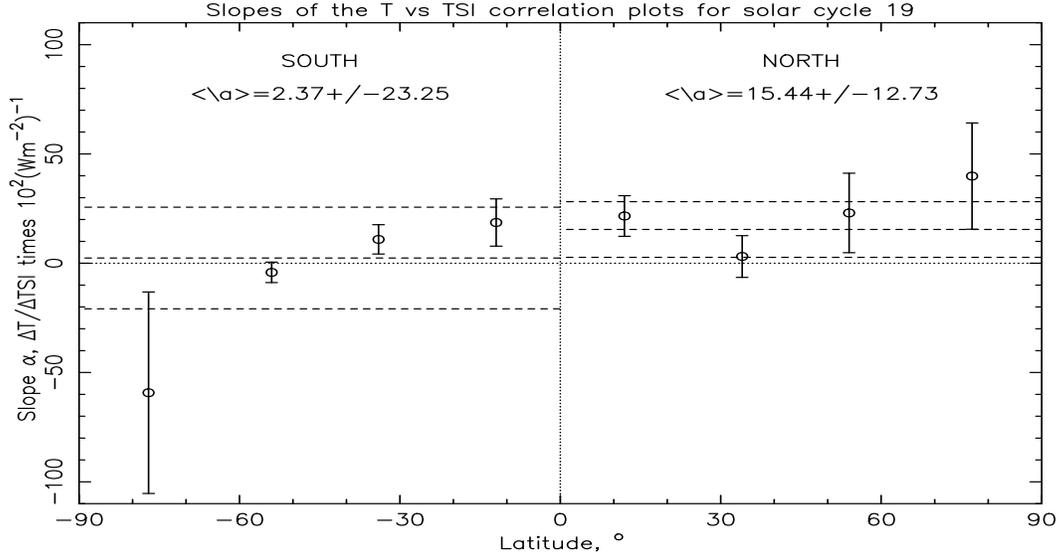}
\end{center}
\caption{\footnotesize $\alpha$-values for Solar Cycle No.19. The vertical lines
represent the means and associated one standard deviation errors. The overall mean
value is $0.06\pm0.13$.}
\label{fig:fig4}
\end{figure}

\subsection{The hemispheric variation of $\alpha$ for single Solar Cycles}
We have averaged the values of $\alpha$ for single Solar Cycles shown in Figure 3
separately for Northern and Southern hemispheres.
Figure 5 shows the results. The dispersion of the values of $\alpha$ about their mean
is clearly much greater for the North than for the South.

The mean values of $\alpha$, which includes the stated errors, give the values
indicated in the Figure. There is an indication of the slow rise of the sensitivity
of northern T to TSI with time ($\alpha=(2.29\pm1.99)\cdot 10^{-2}$ K(Wm$^{-2})^{-1}$ per cycle,
which reverses after 1975 ($\alpha=-(14.3\pm 11.5)\cdot 10^{-2}$ per cycle in the same units),
starting from the 21st Solar Cycle. Again this feature is better seen in the North than in
the South, where $\alpha$ has no time variability.
\begin{figure}[htb]
\begin{center}
\includegraphics[width=16cm,height=12cm]{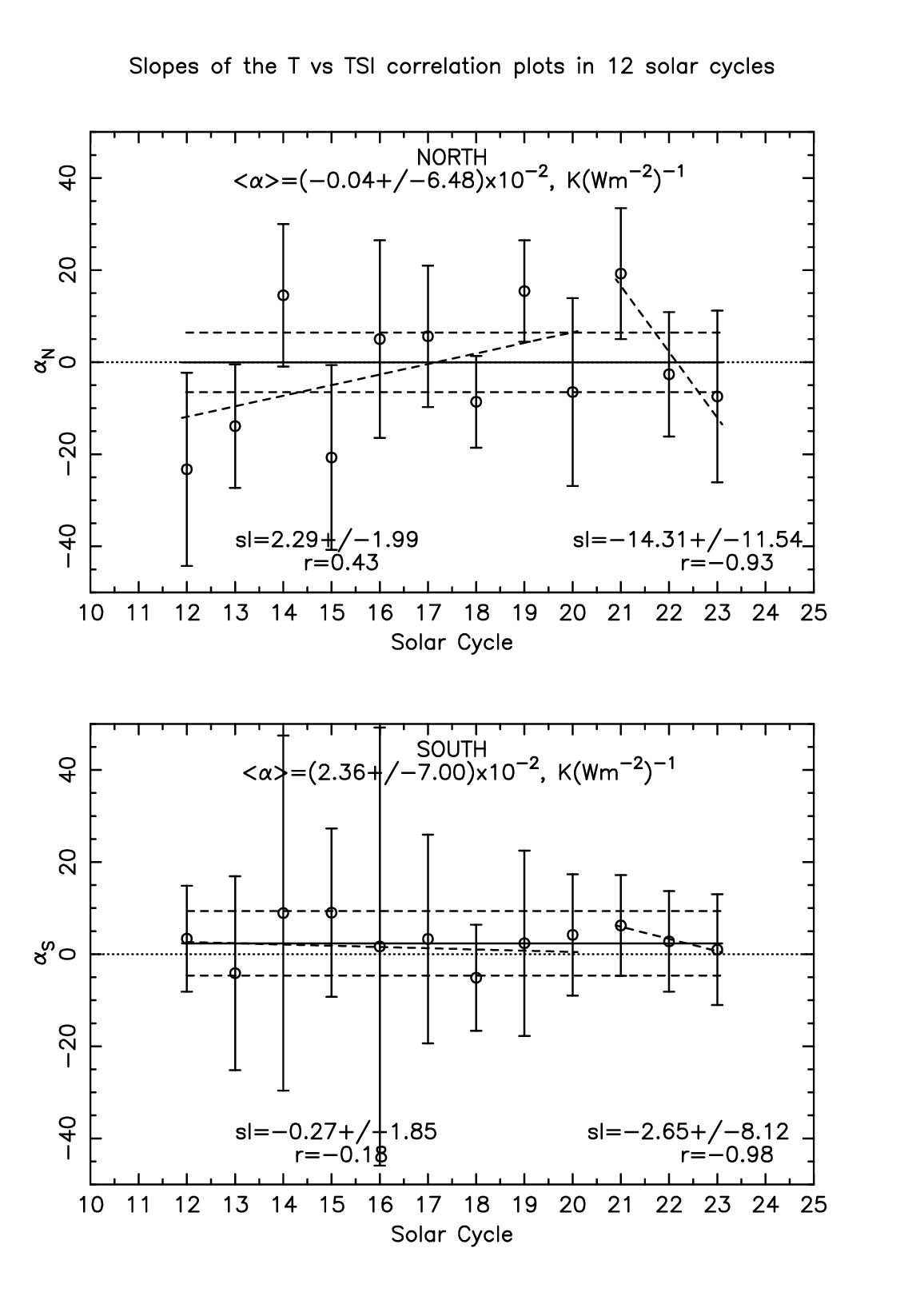}
\end{center}
\caption{\footnotesize The sensitivity ($\alpha$) of the $\Delta$T vs TSI correlation
plots for North and South and for each Solar Cycle, 12-23 inclusive. $\alpha$ is
measured in units of $K(Wm^{-2}$)$^{-1}$. The means and their one standard deviation
errors are indicated.}
\label{fig:fig5}
\end{figure}

\subsection{An alternative method for determining $\alpha$}
Instead of the TSI, a related qualitity, the Sun-spot number (SSN) has been adopted:
the two are not quite exactly proportional. The results are shown in Figure 6. The
corresponding value of $\alpha$ is: $<\alpha> = (0.06 \pm 0.03) K (Wm^{-2}$)$^{-1}$
as described in the caption.
\begin{figure}[h]
\begin{center}
\includegraphics[width=8cm,height=16cm,angle=-90]{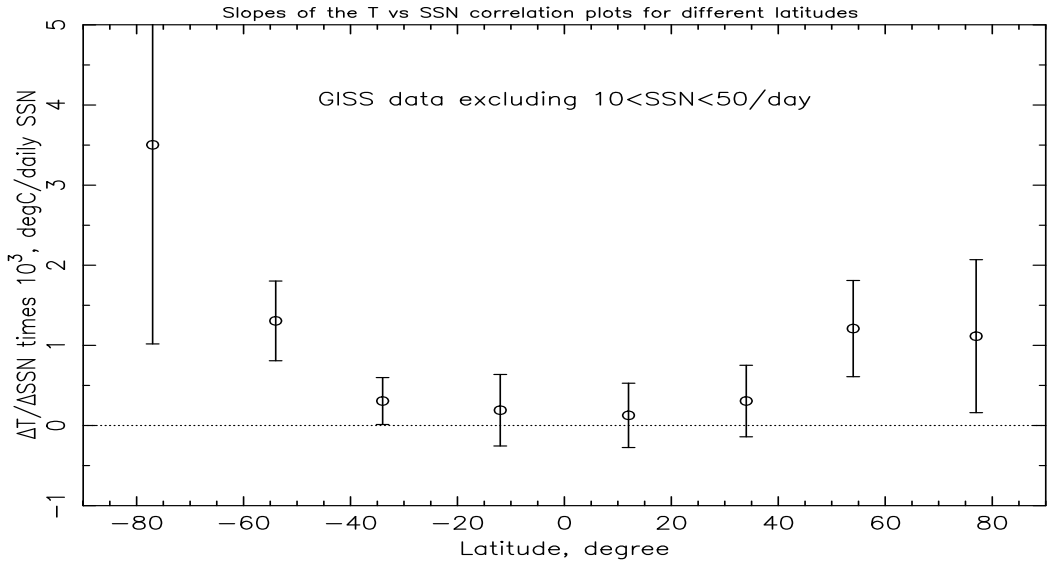}
\end{center}
\caption{\footnotesize Slopes of $\Delta$T versus Sunspot numbers (SSN) for each
latitude range. On this scale, unity corresponds to $\alpha$ =
0.15$^o$K(Wm$^{-2}$)$^{-1}$. The overall mean, $0.48 \pm 0.24$, corresponds to
$<\alpha> = 0.072 \pm 0.036$K(Wm$^{-2}$)$^{-1}$.}
\label{fig:fig6}
\end{figure}

\subsection{Differences in $\alpha$ correlated with the fraction of the Earth's
surface covered by water}
Returning to Figure 3, we draw attention to the straight lines drawn through the
$\alpha$-values, separately, for Solar Cycles 12-20 and 21-23. The border of
1975 is chosen because this is where the temperature starts to increase rapidly in
 most latitude bands and where anthropogenic gases start to increase significantly.

The possible role of seas/oceans can be considered. Figure 7 shows the temporal speed
of rise for $\alpha$, viz. $d\alpha/dN$, where N is the solar cycle number, versus
fraction of the latitude band covered by water, the `ocean fraction, F'. The
statistically significant linear trend of $d\alpha/dN$ with F for Cycles 12-20
shows that $d\alpha/dN$ for any ocean fraction, F, can be found by weighting it by
 linear interpolation of a high value for F=0 and a low value for F=100\%. This
result clearly arises because of the great thermal inertia of the oceans.
\begin{figure}[htb]
\begin{center}
\includegraphics[width=16cm,height=8cm]{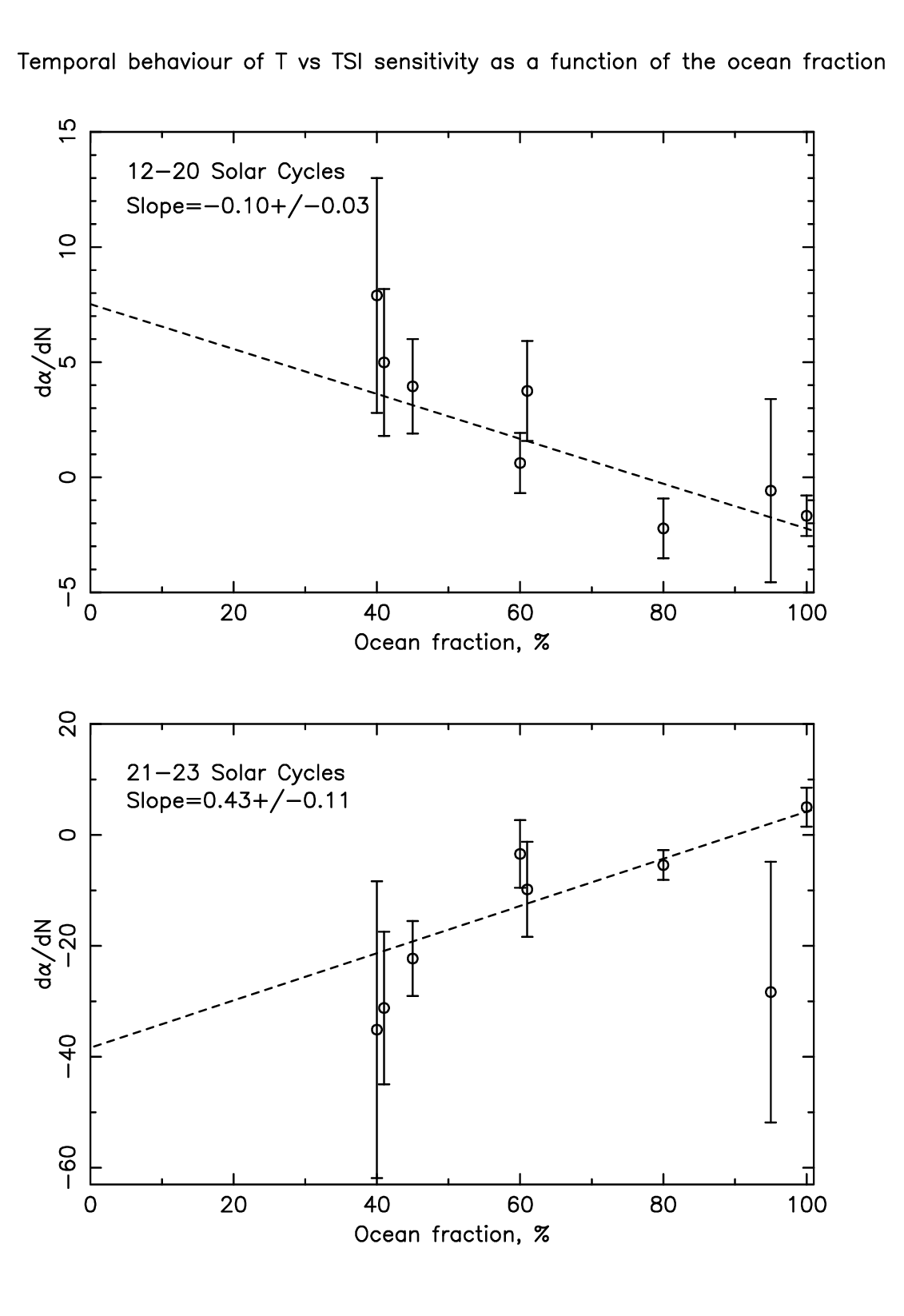}
\end{center}
\caption{\footnotesize Variation of $d\alpha/dN$, where N is the Solar Cycle number, as
 a function of the ocean fraction: upper panel is for Cycles 12-20, lower panel is for
Cycles 21-23.}
\label{fig:fig7}
\end{figure}

The inversion in $d\alpha/dN$ for cycles 21-23, i.e. where the temperature is
increasing rapidly, implies that another mechanism is involved. A likely cause is,
as remarked earlier, the anthropogenic gases which start to rise rapidly at about 1960 
(Etheridge et al. 1996) and thus their delayed effect will start somewhat later.

We have also searched for a correlation of temperature as such with the ocean
fraction, with the results shown in Figure 8.
\begin{figure}[h]
\begin{center}
\includegraphics[width=8cm,height=16cm,angle=-90]{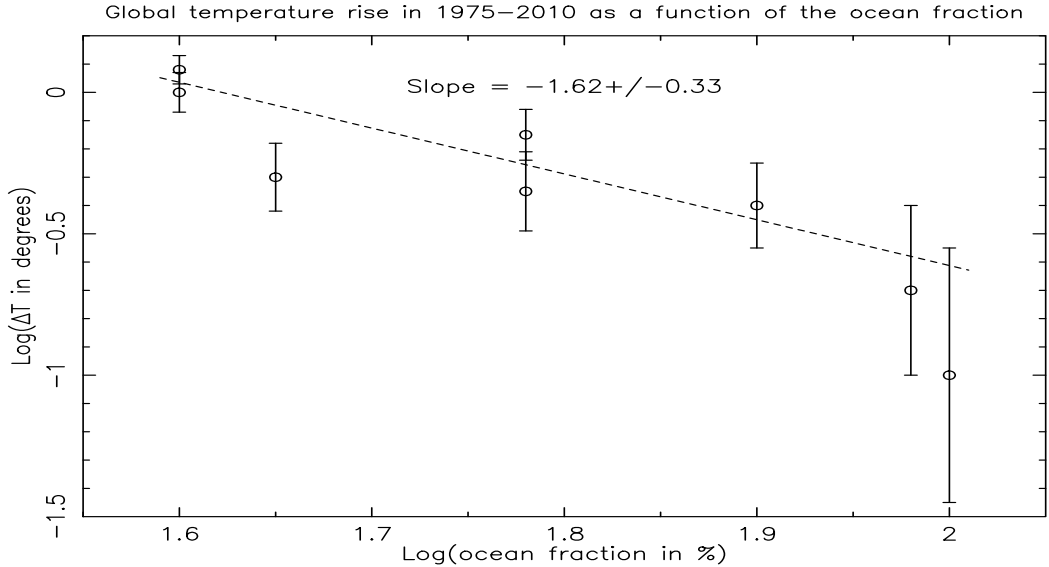}
\end{center}
\caption{\footnotesize Temperature changes from 1975 to 2010, $\Delta$T for each
latitude band as a function of the ocean fraction, F. Note that we plot log$\Delta$T
against log F.}
\label{fig:fig8}
\end{figure}
It is seen that the higher ocean fraction corresponds to the decrease of the
temperature rise observed in 1880-1975. It is another piece of evidence showing the
importance of the higher thermal inertia of the ocean compared with that of the land.

\subsection{Time delays caused by thermal inertia}
It is obvious that thermal inertia will cause a delay in response to a change in energy
 input (TSI) as well as a reduction in the magnitude of $\alpha$. The average TSI has 
an apparent minimum about 1890, which might be expected to manifest itself in a minimum
 in the ensuing temperature change. Using temperature plots for the various latitude
ranges from (TSI 2014) we have determined the year of the first minimum (e.g. 1920 for 
44-64S, where there is nearly 100\% ocean). A plot of delay versus ocean percentage 
yields a straight line at the 3.7 sigma level (2\% probability). For 100\% ocean the 
delay is about 30 years. Similar results appear for other features of the temperature 
series (~Figure 9~). Clearly, such long delays are responsible for the reduced values 
of $\alpha$ for solar cycles, but less so for intervals of several decades.
\begin{figure}[htb]
\begin{center}
\includegraphics[width=16cm,height=12cm]{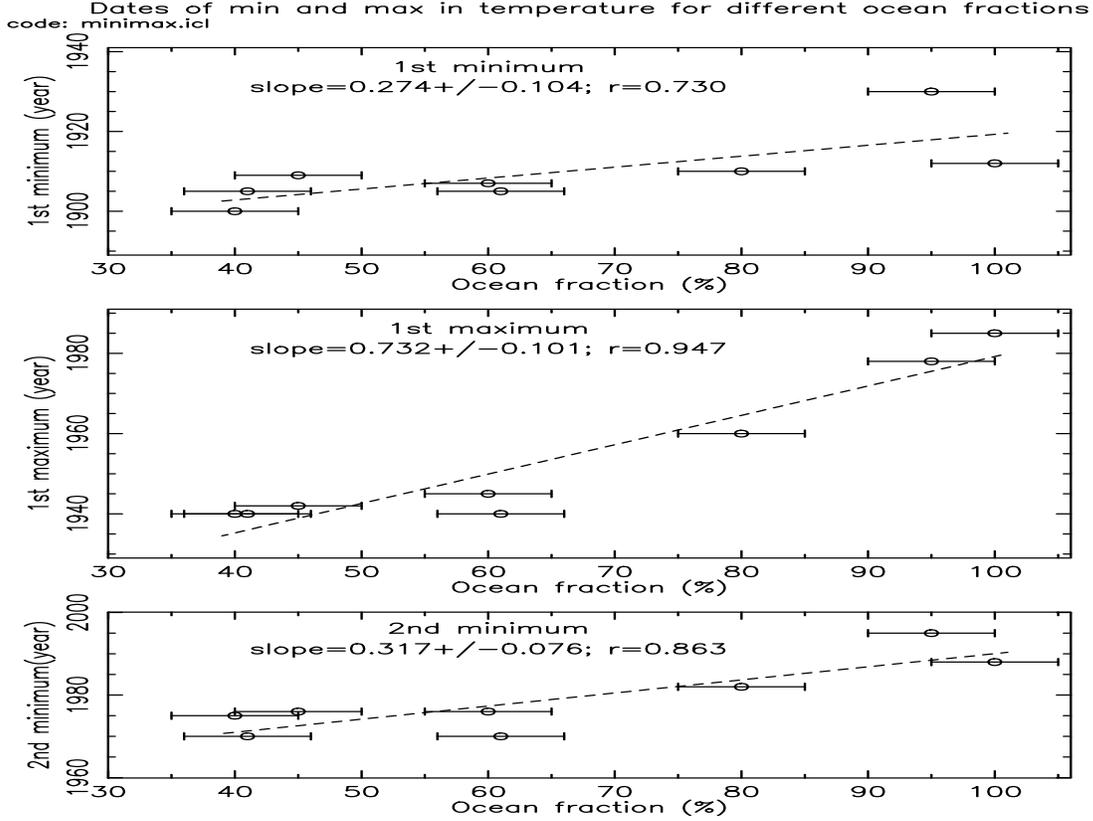}
\end{center}
\caption{\footnotesize Year of the minimum and maximum in temperature profile for
different ocean fractions.}
\label{fig:fig9}
\end{figure}
Returning to Figure 7, if a linear extrapolation of the fit to the lowest ocean
fraction is allowed, the value of $\alpha$ is 0.075 $K(Wm^{-2})^{-1}$ over land alone.

\subsection{Relevance of CO$_2$ to latitudinal variations}
We continue by considering the source of the $CO_2$ density observations. The 
dependence of the global mean $CO_2$ density as a function of time is from (IPCC 
 2014b). Recent values are, in ppm, for the year in brackets: 315(1960), 323(1970), 
335(1980), 352(1992), 366(2000) and 385(2010). There is very little difference between 
the N- and S-hemispheres but there are small latitude variations such that the highest 
$CO_2$ densities are for 35$^\circ$N and 32$^\circ$S, for July, 2008.

There is seen to be some correlation with the results of Figure 2, i.e. the slopes of 
$\alpha$ vs SSN, for 24$^\circ$-44$^\circ$N and 24$^\circ$-44$^\circ$S are slightly 
smaller than the overall average, beyond Solar Cycle 20.

On a longer time scale, it can be noted that there is apparent evidence for a direct 
connection between $\alpha$ and the TSI itself, independently of any $CO_2$ effects. 
This shows itself as a temporal correspondence (1975) between the maximum value of 
$\alpha$ (Figure 10) and that of the TSI (Figure 1). 

Explanations will be advanced later.

 Following the remarks in the previous sections about anthropogenic gases 
(referred to here as `CO$_2$'), we give, in Figure 10, a plot of $\alpha$ as such 
against $CO_2$ content with respect to a datum of 240ppm (IPCC 2014b).
\subsection{The average value of $\alpha$ for time ranges of several decades}
Figure 2 gives an average (for N and S) over nearly 95 years of
$(17 \pm 3)\cdot 10^{-2}$ $K(Wm^{-2}$)$^{-1}$ and in the text (\S2.2). For a longer
period, some 350y, we have $(21 \pm 5)\cdot 10^{-2}$ $K(Wm^{-2}$)$^{-1}$.
Both are shown in Figure 11 and both are seen to be similar to other estimates (by us,
Erlykin and Wolfendale 2010).
\subsection{An estimate of $\alpha$ for the last 6000 years}
Inevitably, measurements of Temperature and TSI over historic time (thousands of years)
are inaccurate, being derived from proxies. However, tentative estimates have been made
 (Mann et al. 1998) from which an approximate value of $\alpha$ can be derived. No 
doubt other parameters affect the global temperature besides TSI but there is a 
tolerable correlation of $\Delta T$ and (proxy) sunspot number for the last 6000 years.
 The approximate value of $\alpha$ is 0.29 $K(Wm^{-2})^{-1}$.
\subsection{The effect of ENSO and PDO on $\alpha$}
In the Introduction it was pointed out that ENSO and PDO were ignored, i.e. their 
effect on the derivation of 
$\alpha$ was not included. It is of interest, however, to see to what extent $\alpha$
is dependent on ENSO and PDO, and other parameters, too, specifically the number of
sunspots at maximum, the 'width' of the cycle ( i.e. minimum to minimum ) and the aa
magnetic index. In each case $\alpha$ is plotted against the parameter in question and
a straight line fitted. The slopes of these lines in terms of their standard deviation
from the mean are given in Table 1 for both the North and South hemispheres.
\begin{center}
\begin{tabular}{||c|c|c||}
\hline\hline
\bf{Parameter}          & {\bf North} & {\bf South} \\       \hline\hline
     PDO                & 2.6         & 0.7         \\       \hline
     ENSO               & 0.3         & 2.0         \\       \hline
     Max.SSN            & 1.4         & 2.0         \\       \hline
     Cycle Width        & 4.4         & 5.2         \\       \hline
     aa magnetic index  & 4.4         & 0.2         \\       \hline\hline
\end{tabular}
\end{center}
{\footnotesize Table 1. Slopes of $\alpha$ versus parameter in terms of their number of
 standard deviations from zero}

It will be noted that the only (statistically) significant correlations are with cycle
width and the aa magnetic index, both indicators directly related to the Sun.
\begin{figure}[htb]
\begin{center}
\includegraphics[width=16cm,height=12cm]{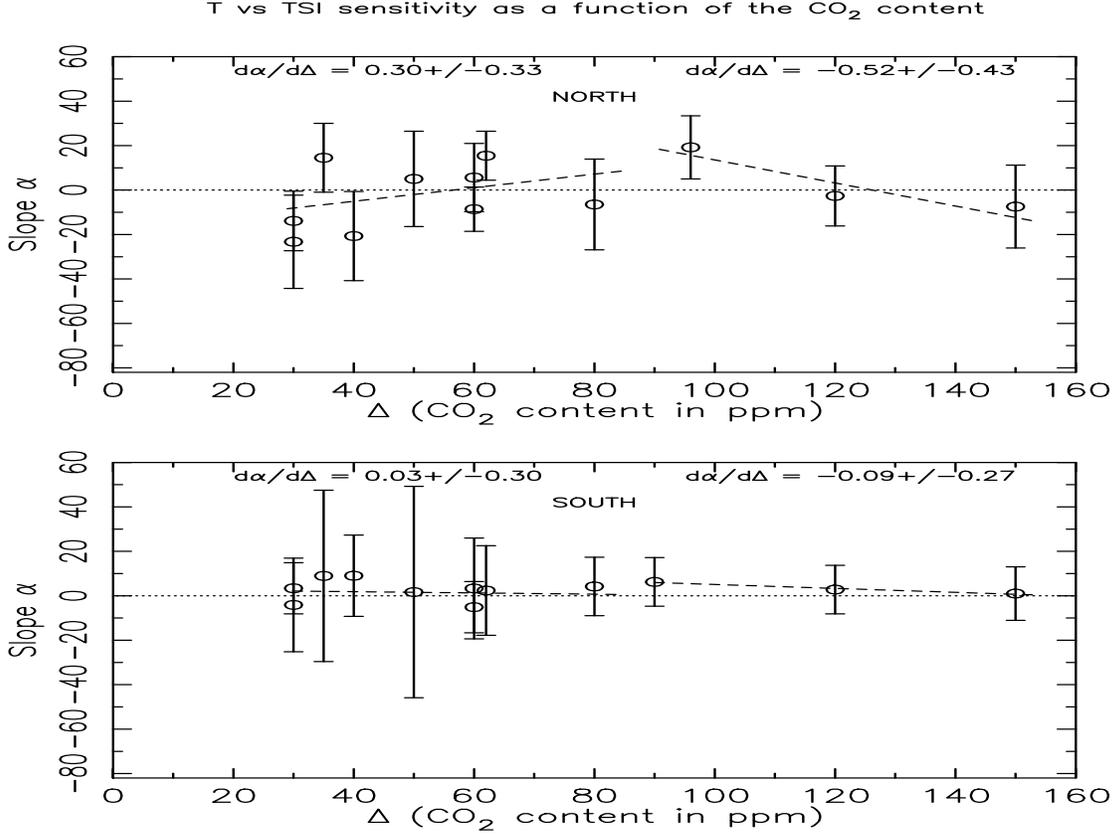}
\end{center}
\caption{\footnotesize Temporal behaviour of $\alpha$ as a function of CO$_2$ content,
the $CO_2$ datum being 240ppm. 90ppm corresponds approximately to the situation for 
1975. It is evident that there must be additional sources of error in the North for 
$\Delta \leq$ 100}
\label{fig:fig10}
\end{figure}
\section{Discussion of the Results}
\subsection{The validity of the $\alpha$-values and their explanation}
There is evidence for a finite but very small value of $\alpha$ for Solar Cycles, as
such. An average estimate from the results presented is
$<\alpha> = (0.07 \pm 0.02)K(Wm^{-2})$, averaged from 1890 using the values from
Figures 4, 6 and 7. This is indicated in Figure 11, where it is compared with the
results of other workers: it is seen to be in the region where there is rough 
consistency between the results of others.

There seems little latitudinal variation of the values of $\alpha$ apart from the clear
 correlation with ocean fraction (see Figure 7) and the small effect mentioned in 
\S3.6. The increasing diminution with time after Cycle 20, which is a prominent feature
 in Figure 3, is attributed, by us, to the effect of anthropogenic gases, as remarked 
already, but, in fact the correlation in Figure 10 needs further study (see \S4.3).
\begin{figure}[htb]
\begin{center}
\includegraphics[width=8cm,height=16cm,angle=-90]{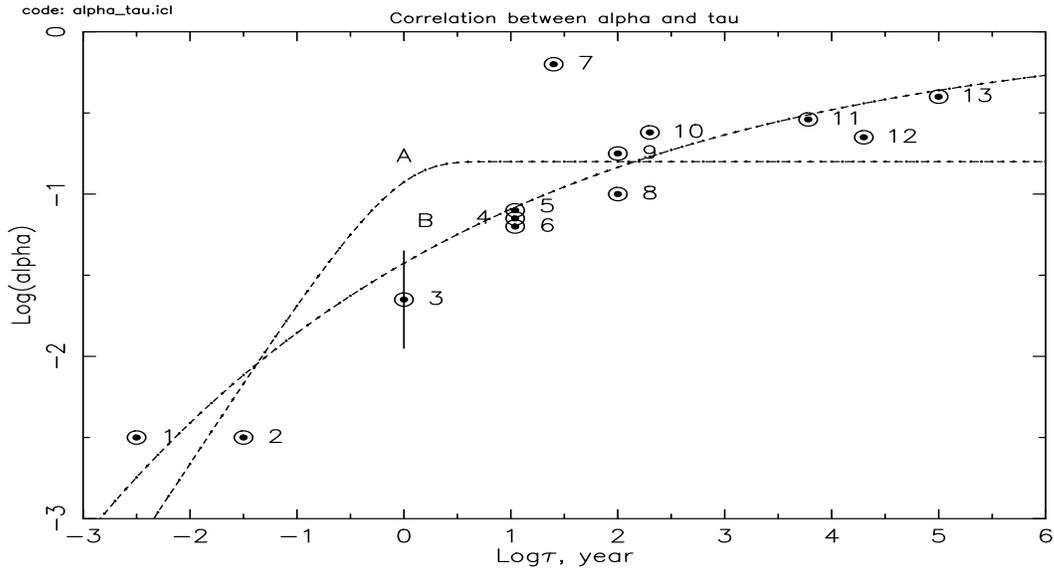}
\end{center}
\caption{\footnotesize $\alpha$-values, i.e. sensitivity of mean global temperature to
Total Solar Irradiance, over various time intervals $\tau$, $\alpha$ is measured in
$K(Wm^{-2}$)$^{-1}$. Key: see separate sheet. Note the logarithmic scales.}
\label{fig:fig11}
\end{figure}

\subsection{The time dependence of $\alpha$}
Figure 7 represents an apparently new feature of the interrelation of the rate of
change of $\alpha$ with ocean fraction and some discussion is necessary.

Firstly, the value of $\Delta\alpha/\Delta N \sim 0$ for 100\% ocean fraction in both
cases (TSI rising and TSI falling) is interesting, in that it shows that this situation
 (`all ocean') leads to a near temporal stability of $\alpha$.

The difference in the slopes before and after 1975 indicates that, for the `mainly
land' region, $\Delta\alpha/\Delta N$ is more sensitive to TSI changes after 1975 than
before. It is perhaps the effect of TSI on the CO$_2$ which is largely responsible for 
the `greenhouse effect' or simply the reduction in $\alpha$ for the 11-year cycle 
caused by the fact that there is no (appreciable) 11-year cyclical variation in the 
$CO_2$ density, i.e. the temperature goes up but TSI cyclical changes have a 
diminishing effect. The latter explanation is preferred.

The extrapolation for 'all-land' is seen to be high - as would have been expected, and
may be relevant for large land areas with comparatively stable climates.

\subsection{The latitude dependence of $\alpha$}
It was mentioned at the start that there might be additional information, related to 
climate change, from analysis of the latitude variation of various parameters.

That this is the case can be seen from Figures 4, 6 and 7. The `latitude-effect' 
appears to be well-explained by the dependence of the ocean-fraction on latitude, the 
increased thermal inertia of the oceans having a damping effect which reduces $\alpha$.
 The damping period can be seen in Figure 9.

The important role of 1975 as the year when the `greenhouse effect' due largely to 
$CO_2$ started to dominate the temperature rise (Figure 8.)

An interesting further fact can be seen in Figure 3, where for Cycles 21, 22 and 23, 
the slopes vary with latitude in a particular way. Figure 8 gives an explanation in 
terms of oceanic fractions. Inspection of IPCC Report (IPCC 2014b) reveals estimates of
 the likely temperature rises as a function of latitude and longitude, for the period 
2016 to 2036, $CO_2$ being largely responsible for the temperature increases. Their 
latitude dependence mirrors that of the negative slope of $\alpha$ versus Solar Cycle 
No, N, as can be seen in Table 2. Thus, the higher the efficiency of T-change due to 
$CO_2$, ($\Delta T(CO_2))$, the greater the (negative) slope of $\delta \alpha / 
\delta N$.

\begin{center}
\begin{tabular}{| l | c | r |}
\hline
Latitude Range (deg.)   & $\delta\alpha/\delta N$   & $\Delta$T(CO$_2$) \\ \hline
$>$64$^o$N              & 1.2                       & 2.0 \\ \hline
44-64$^o$N              & 1.0                       & 1.4 \\ \hline
24-44$^o$N              & 0.35                      & 0.9 \\ \hline
0-24$^o$N               & 0.8                       & 0.9 \\ \hline
0-24$^o$S               & 0.45                      & 0.5 \\ \hline
24-44$^o$S              & 0.25                      & -0.1 \\ \hline
44-64$^o$S              & -0.25                     & 0 \\ \hline
$>$64$^oS$              & 0.8                       & 0.7 \\ \hline
\end{tabular}
\end{center}

{\footnotesize Table 2. Latitude variation of the variation of $\alpha$ with Solar 
Cycle No. ($\delta \alpha / \delta N$), in arbitrary units, (from Figure 3) and the 
predicted change in temperature from 2016 to 2036 ($\Delta T(CO_2))$. in deg. K (IPCC 
2014b)}

The fit is at the 5 sigma level and the probability (of non-correlation) is 0.5\%.

The fraction of the `recent' temperature rise due to changes in the TSI is an important
 parameter that needs discussion. If we consider 1900 as the starting point and 1975 as
 a recent datum, the appropriate value of $\alpha$ to use is 0.2$K(Wm^{-2})^{-1}$
(Figure 11). The increase in temperature expected for $\Delta$(TSI) of
0.6Wm$^{-2}$ is thus 0.12$^o$K. This can be compared with the measured increase of
0.4$^o$K (TSI 2014), i.e. 30\%. This is, clearly, not a negligible fraction.

Since 1975 the TSI has fallen by 0.46$(Wm^{-2})^{-1}$ and our expected temperature
reduction is 0.10$^o$K. In this time the mean global temperature rise has been 0.50
$^o$K thus the TSI change gives a reduction of 20\%. Without the TSI change, the
temperature rise would have been 0.6$^o$K. The 0.10$^o$K difference is not negligible
and may contribute to the apparent fall in rate of rise in temperature after 1975 (the
well-known `hiatus').

\section{Comparison with the results of other workers and conclusions}
The results are compared in Figure 11. The values for 1 day and 1 year are simplistic
determinations, by us, for day/night temperature differences and seasonal differences,
respectively. There is a general consensus of the values of $\alpha$.

Considering the 11-year solar cycle, it is interesting to note that those 
determinations which studied the individual components of temperature change: TSI, 
ENSO, volcanoes and a general rise (Lockwood 2007; Lean 2004; Haigh 2007)
 denoted 5 and 6, reached the same `solar value' as our mean: (4) in Figure 11. This is
 not to say that ENSO et al. have no effect on climate but their effects come 
partly from TSI. This is also true to some extent for CO$_2$ changes which are, in 
part, caused by increased ocean temperatures.

The global values of $\alpha$ are collected together in Figure 9 and details are given 
in Table 2. Smooth lines through the points illustrate the effect of the thermal
inertia of the globe. The lines need an explanation. That marked A is the standard form
 for, for example, the charging of a condenser through a resistor:
 $\alpha = \alpha_0 (1-exp(-\frac{\tau}{\tau_0}))$ where $\tau_0$ is the effective time
 constant. That marked B is for the standard form but for $log \alpha$ versus
$log \tau$ and is thus more of an empirical fit.

A possible explanation is that we have the sum of two (weighted) curves of type A, one
for land and one for oceans, the values of $\tau_0$ being different. In first
approximation $\tau_0 \simeq 1$ year for land and $\simeq 30$ years for ocean.

To conclude, we have demonstrated that solar effects become increasingly important
until about 1975, following which another mechanism (anthropogenic gases?) take over.
Understandably, the role of the oceans, and their considerable thermal inertia, is
important.

There is consistency between the results from different authors and a smooth plot of
sensitivity of atmospheric temperature versus change in TSI seems appropriate from days
 to hundreds of thousands of years.

$\bullet$ All the factors considered by us confirm the conclusion by climatologists 
that TSI changes dominated as the cause of increasing global temperatures before about 
1975 but that another cause (presumably anthropogenic gases) is responsible for the 
`rapidly' increasing temperatures in the most recent 40 years.

{\bf Acknowledgements} \\
The authors are grateful to the Kohn Foundation for financial support and to Professor 
Terry Sloan for helpful comments.

\newpage

\section{Table 3. Key to the points in Figure 11}
Slopes ($\alpha$-values) from various sources versus the 'integration time' to which
the values relate ($\tau$). Typical errors are $\pm$ a factor 2.
\begin{enumerate}
\item Daily variation. Present work. 1d.
\item Forbush decreases. $\sim$10 d. (Laken 2012)
\item Summer/Winter vs. latitude. 1 year. (Erlykin and Wolfendale 2010)
\item Solar Cycle. Present work: \S4.1.       11 years
\item Solar Cycle.                      11 years (Lockwood 2007, Lean et al. 2005)
\item Solar Cycle.                             11 years (Haigh 2007) 
\item 'Long dip'.                              25 years (Erlykin and Wolfendale 2010)
\item Maunder minimum.                        100 years (Lean et al. 2005)
\item Previous decades. Present work: \S4.2.  100 years
\item Ditto.                                  200 years
\item Long-term. Present work: \S2.4.        6000 years
\item Last Ice Age.                  $2\cdot10^4$ years (Haigh 2007)
\item Orbital changes.                     $10^5$ years. Rusov V. et al., 
2008, private communication (siiis@t.e.net.ua)
\end{enumerate}

{\large{\bf References}}

\vspace{2mm}

Erlykin AD, Gyalai G, Kudela K, Sloan T, Wolfendale AW, 2009, Some aspects 
of ionization and the cloud cover, cosmic ray correlation problem, J. Atmos. and 
Solar-Terr. Physics,71, 823-829; doi:10.1016/j.jastp.2009.03.007 \\

Erlykin AD, Wolfendale AW, 2010, Global cloud cover and the Earth's mean surface 
temperature, J. Atmos. and Solar-Terr. Physics, 31/4, 399-408; doi:10.1007/
s10712-010-9098-7(10pp)\\

Etheridge DM, Steele LP, Langenfelds RL, Francey RJ, Barnola JM,  Morgan I, 1996, 
Natural and anthropogenic changes in atmospheric $CO_2$ over the last 1000 years from
 air in Antarctic ice and firn, J. Geophys. Res., 101, 4115-4128 \\

Foukal P, 2012, A new look at solar irradiance variation, Solar Physics, 279, 365-381; 
doi:10.1007/s11207-012-0017-6(17pp) \\

Haigh J.D., 2007, The Sun and the Earth's climate, Living Rev. Solar Phys., 4, 2 \\

IPCC Report, 2014a, Climate change, the physical science basis, ch 8 Anthropogenic and 
natural radiation forcing, 659-740 \\

IPCC Report, 2014b, Climate change, the physical science basis, ch 11 Near-time climate
 change projections and probability, 959-1029 \\

Laken B , Wolfendale AW, Kniveton D, 2012, Cosmic ray decreases and changes in the 
liquid water fraction over the oceans,  Geophys. Res. Lett., L23803(5pp) \\

Lean JL., 2004, Solar irradiance reconstruction, IGBP pages/World Data Centre for 
Paleoclimatology, Data contribution series \#2004-035, NOAA/NGDC Paleoclimatology 
Program, Boulder, Co, USA (29pp) \\

Lean JL, Rottman G, Harder J, Kopp G, 2005, SORCE contribution to new understanding to
global change and solar variability, Solar Phys., 230, 27-53 \\

Lockwood M, 2007, Recent changes in solar outputs and the global mean solar 
temperature. III Analysis of contributions to global mean air surface temperature rise.
 Proc. Roy. Soc. A, doi:10.1098/rspa.2007.0348 \\

Mann ME, Bradley RS, Hughes MK, 1998, Global-scale temperature patterns and climate forcing over the past six centuries, Nature, 392, 779-787 \\

Marsh ND, Svensmark H, 2000, Low cloud properties influenced by cosmic rays, Phys. Rev.
 Lett., 85, 5004-7 \\

Scafetta N, West B.J., 2007, Phenomenological reconstructions of the solar signature in
 the Northern Hemisphere surface temperature records since 1660, J. Geophys. Res., 112,
 D24503 (10pp) \\

TSI 2014,  $http://lasp.colorado.edu/data/sorce/tsi\_data$ \\

Wang Y-M, Lean JL, Speeley NR, 2005, Modeling the Sun's magnetic field and irradiance 
since 1713, Astrophys. J., 625, 522-538 \\

Zacharias P, 2014, An independent review of existing total solar irradiance, Surveys in
 Geophys., 35, 897-912 \\

Zonal temperatures: $http://data.giss.nasa.gov/gistemp/tabledata\_v3$

\end{document}